\documentclass[onecolumn,showpacs,preprintnumbers,amsmath,amssymb]{revtex4}
\usepackage{graphicx}
\usepackage{dcolumn}
\usepackage{bm}
\usepackage{subfigure}
\usepackage{marvosym}
\usepackage{textcomp}
\usepackage{threeparttable}
\usepackage{booktabs}
\usepackage{setspace}
\usepackage{type1cm}
\usepackage{color}
\renewcommand{\TPTtagStyle}%
{\normalsize\textit}

\begin{document}
\fontsize{12pt}{12pt}\selectfont

\preprint{UTTG-11-12}

\title{Jet Quenching and Holographic Thermalization with a Chemical Potential}
\author{Elena Caceres$^{1,2}$, Arnab Kundu$^2$, Di-Lun Yang$^3$}
\affiliation{$^1$Facultad de Ciencias, Universidad de Colima, Bernal Diaz del Castillo 340, Colima, Mexico. \\
$^2$Theory Group, Department of Physics,
University of Texas at Austin, Austin, TX 78712, USA.\\ 
$^3$Department of Physics, Duke University, Durham, North Carolina 27708, USA.}

\begin{abstract}
We investigate jet quenching of virtual gluons and thermalization of a strongly-coupled plasma with a non-zero chemical potential via the gauge/gravity duality. By tracking a charged shell falling in an asymptotic AdS$_{d+1}$ background for $d=3$ and $d=4$, which is characterized by the AdS-Reissner-Nordstr\"om-Vaidya (AdS-RN-Vaidya) geometry, we extract a thermalization time of the medium with a non-zero chemical potential. In addition, we study the falling string as the holographic dual of a virtual gluon in the AdS-RN-Vaidya spacetime. The stopping distance of the massless particle representing the tip of the falling string in such a spacetime could reveal the jet quenching of an energetic light probe traversing the medium in the presence of a chemical potential. We find that the stopping distance decreases when the chemical potential is increased in both AdS-RN and AdS-RN-Vaidya spacetimes, which correspond to the thermalized and thermalizing media respectively. Moreover, we find that the soft gluon with an energy comparable to the thermalization temperature and chemical potential in the medium travels further in the non-equilibrium plasma.  
The thermalization time  obtained here by tracking a falling  charged shell does not exhibit, generically,  the same qualitative features as the one obtained studying non-local observables. This indicates that  --holographically--  the definition of thermalization time is  observer dependent and there is no  unambiguos definition. 
\end{abstract}

\newpage

\maketitle

\section{\label{sec:level1}Introduction}

The AdS/CFT correspondence, a duality between the type IIB supergravity and the $\mathcal{N}=4$ $SU(N_c)$ super Yang-Mills theory (SYM) at large 't Hooft coupling in the large $N_c$ limit\cite{Aharony:1999ti,Gubser:1998bc,Witten:1998qj,Maldacena:1997re}, has been widely studied to investigate the properties of strongly coupled systems. At finite temperature, the spacetime metric in the gravity dual is governed by the AdS-Schwarzschild geometry, in which the temperature is characterized by the Hawking temperature of a black hole\cite{Witten:1998zw}. The holographic dual description can be regarded as an analogue of the strongly-coupled quark gluon plasma (QGP) generated in the relativistic heavy ion collisions. On the other hand, the QGP also carries a non-vanishing chemical potential. In the gravity dual, the chemical potential is encoded in the AdS-Reissner-Nordstr\"om (AdS-RN) spacetime\cite{Chamblin:1999tk,Cvetic:1999ne} which carries a non-vanishing gauge field.

One of the interesting aspects of the heavy ion physics is the thermalization process of the medium after the collisions of the two nuclei. In the gravity dual, this scenario should correspond to the gravitational collapse and the formation of a black hole. The issue was studied by implementing various approaches in the literature. In \cite{Grumiller:2008va,Gubser:2008pc,Albacete:2008vs,AlvarezGaume:2008fx,Lin:2009pn,Albacete:2009ji,Gubser:2009sx,Kovchegov:2009du,Aref'eva:2009kw,Kovchegov:2010zg,Taliotis:2010pi,Chesler:2010bi,Kiritsis:2011yn,Taliotis:2012sx,Aref'eva:2012ar,Casalderrey-Solana:2013aba} the collisions of gravitational shock waves are introduced to mimic the colliding nuclei in the relativistic collisions. In \cite{Heller:2011ju,Heller:2012je} time-dependent and boost-invariant metrics, associated with the plasma undergoing Bjorken expansion, were investigated and further generalized in \cite{Kalaydzhyan:2010iv} with a chemical potential. The isotropization time in such a framework was firstly estimated in\cite{Kovchegov:2007pq}. Moreover, the authors in \cite{Chesler:2008hg,Chesler:2009cy} introduced anisotropic and time-dependent boundary conditions, which lead to the thermalization of an anisotropic plasma. Similar studies can be found in \cite{Heller:2012km,Heller:2013oxa}. Alternatively, the gravitational collapse can be characterized by a collapsing shell, which results in the isotropic thermalization\cite{Giddings:2001ii,Lin:2008rw,Bhattacharyya:2009uu,Garfinkle:2011hm,Garfinkle:2011tc,Erdmenger:2012xu,Wu:2012rib,Wu:2013qi}.
The collapse of an inhomogeneous shell was further investigated in \cite{Balasubramanian:2013rva,Balasubramanian:2013oga}.  
Note that, such {\it first principle} computations are generally difficult and thus a more {\it phenomenological} approach allows us to probe the physics with more ease.

We also note that the background metric generated by a time-varying weak dilaton field produces AdS-Vaidya-type background\cite{Bhattacharyya:2009uu} and provides a very good quantitative approximation in the thin-shell limit. In \cite{Balasubramanian:2010ce,Balasubramanian:2011ur,Arefeva:2012jp}, various non-local operators have been studied in the AdS-Vaidya spacetime to probe  aspects of thermalization of the medium; and generalized in the presence of a charged shell where the spacetime is represented by AdS-RN-Vaidya metric\cite{Caceres:2012em,Galante:2012pv}. As shown in \cite{yang}, a thermalization time which is independent of the length scales of non-local operators can be extracted from an alternative approach.

Another salient issue in the heavy ion collisions is the jet quenching of hard probes traversing the medium. Much effort has been made towards studying this phenomenon in the thermalized medium {\it via} holographic methods. In general, the heavy probes such as heavy quarks are assumed to constantly travel through the infinite medium. In \cite{Gubser:2006bz,Herzog:2006gh}, jet quenching of heavy quarks is characterized by the drag force caused by the interaction with the medium, which can be derived from the dynamics of a trailing string in the gravity dual. Furthermore, the jet quenching parameter, which encodes the momentum broadening of hard probes, can also be extracted from the computation of the light cone Wilson loop in the curved spacetime\cite{Liu:2006he,Liu:2006ug,D'Eramo:2010ak}. However, for light probes such as light quarks and gluons, they may finally dissipate in the medium. According to \cite{Hatta:2008tx,Gubser:2008as,Chesler:2008uy,Arnold:2010ir,Arnold:2011qi,Arnold:2012uc,Muller:2012uu}, the gravity dual of light probes may have various candidates. The maximum stopping distance of the light probes can, nevertheless, be derived from the null geodesic of a massless particle falling in the dual geometry. There are only a handful of studies on the influence of a non-equilibrium medium on the jet quenching of hard probes\cite{Spillane:2011yf,Stoffers:2011fx,Chesler:2013cqa,yang}. Some work on the influence of thermalization on electromagentic probes can be found in \cite{Baier:2002tc,Baier:2012tc,Steineder:2012si}. In \cite{yang}, it is indicated that jet quenching of light probes with energy much greater than the temperature of the thermal medium remains unaffected by the non-equilibrium processes. This situation may change when the probes carry finite energy comparable with other soft scales of the medium.

In this paper, we will follow the previous studies in \cite{Caceres:2012em,yang} to investigate thermalization of the medium with a chemical potential and jet quenching of light probes in such a non-equilibrium plasma. Our work is organized in the following order. In section II, we analyze the AdS-RN-Vaidya metric in Poincare coordinate and extract a thermalization time by tracking the position of the falling shell in the thin-shell limit. We compare our results with the ones obtained from non-local observables \cite{Caceres:2012em} and argue that these two prescriptions can only be compared when the length scale of the non-local operators is roughly the size of the future horizon.
In section III following \cite{Gubser:2008as}, we study jet quenching of a virtual gluon traveling in the medium, which is characterized by a double string falling in the gravity dual background. First we compute the stopping distance of a gluon in the thermalized medium with non-zero chemical potential. Then we set up the initial conditions of the falling string in the AdS-RN-Vaidya geometry and the matching condition when the tip of the string penetrates the shell. Towards the end of the section, we evaluate stopping distances for both hard and soft gluons in the non-equilibrium medium with a chemical potential. 
We conclude with a brief summary and discussions. Some technical details are presented in two appendices.

\section{\label{sec:level1}Falling Shell in AdS-RN-Vaidya Spacetime}

In this section, we will extract the thermalization time of the non-equilibrium plasma with a non-zero chemical potential by tracking the position of the falling shell in Poincare patch of AdS-RN-Vaidya geometry. The approach is different from studying non-local operators in Eddington-Finkelstein (EF) coordinates as carried out in \cite{Caceres:2012em,Galante:2012pv}. Since we can define thermalization time to be the time when the shell almost coincides with the future horizon, it is independent of the length of the operators. We will observe that with increasing chemical potential, this thermalization time decreases.

We begin with the AdS-RN-Vaidya\footnote{For $d=3$, the Hodge dual of the bulk electro-magnetic field is another two-form. Thus we can also introduce a magnetic charge in the AdS-Vaidya type background: such a background should be dubbed as AdS-dyon-Vaidya background. By adjusting the values of the electric and magnetic charge in the system, the final temperature of the medium can be fine-tuned to zero, which corresponds to an electro-magnetic quench phenomenon. More discussion on the dyonic background can be found in Appendix B.} metrics in EF coordinates with $d=3$ and $d=4$, respectively. For $d=3$, we have
\begin{eqnarray}\label{efvaidya4d}
ds^2=\frac{1}{z^2}\left(-f(v,z)dv^2-2dvdz+dx_i^2\right) \ ,~~A_v=q(v)(z_h-z) \ , 
\end{eqnarray}
where $f(v,z)=1-m(v)z^3+\frac{1}{2}q(v)^2z^4$. For $d=4$, we have
\begin{eqnarray}\label{efvaidya5d}
ds^2&=&\frac{1}{z^2}\left(-f(v,z)dv^2-2dvdz+dx_i^2\right) \ ,~~A_v=q(v)(z_h^2-z^2) \ ,
\end{eqnarray}
where $f(v,z)=1-m(v)z^4+\frac{2}{3}q(v)^2z^6$. Here we have set the AdS curvature radius $L=1$. The coordinates $x_i$ represent the $d$-dimensional spatial directions. Also, $v$ denotes the EF time coordinate and $z$ denotes the radial direction. The boundary is located at $z=0$ and $z_h$ denotes the future horizon. In the equations above, $m(v)$ and $q(v)$ represent the mass and electric charge of the shell. The charge is related to the time component of the vector potential, which generates an R-charge chemical potential in the gauge theory side {\it via}
\begin{eqnarray}
\mu=\lim_{z\rightarrow 0}A_v(v,z) \ .
\end{eqnarray}
For simplicity, we are interested in the thin-shell limit of the interpolating mass function
\begin{eqnarray}\label{mass}
m(v)=\frac{M}{2}\left(1+{\rm tanh}\left(\frac{v}{v_0}\right)\right)
\end{eqnarray}
with $v_0\rightarrow 0$. We can make the same {\it choices} as \cite{Caceres:2012em} by taking $q(v)^2=Q^2m(v)^{4/3}$ and $q(v)^2=Q^2m(v)^{3/2}$ for $d=3$ and $d=4$, respectively. The values of $z_h$ are determined by
\begin{eqnarray}\label{blackeningfun}
f(v>v_0,z_h)&=&1-Mz_h^3+\frac{1}{2}M^{4/3}Q^2z_h^4=0 \ ,~~\mbox{for}~d=3 \ ,\\  
f(v>v_0,z_h)&=&1-Mz_h^4+\frac{2}{3}M^{3/2}Q^2z_h^6=0 \ ,~~\mbox{for}~d=4 \ .
\end{eqnarray} 
We assume the shell falls from the boundary at $t=0$ and thus $\mu$ is a constant since $v=t\geq 0$ on the boundary. On the other hand, the thermalization temperature is given by
\begin{eqnarray}\label{temperature}
T=-\frac{1}{4\pi}\frac{d}{dz}f(v>v_0,z)|_{z_h} \ .
\end{eqnarray}
The entropy can be determined by the area of the black hole,
\begin{eqnarray}\label{entropy}
S=\frac{A_h}{4\pi G} \ ,
\end{eqnarray}
where $A_h$ represents the area of the black hole and $G$ denotes the Newtonian constant.

Before proceeding further, a few comments are in order. Since the underlying theory is conformal, the only relevant parameter is the ratio $T/\mu$. Thus we define
\begin{eqnarray}
\chi_{d} = \frac{1}{4\pi} \left(\frac{\mu}{T} \right)  \ ,
\end{eqnarray}
which we will consider throughout the paper.

Now, we would like to convert the metric into Poincare coordinates. By taking 
\begin{eqnarray}\label{dv}
dv=g(t,z)dt-\frac{1}{f(v,z)} dz \ ,
\end{eqnarray}
where $g(t,z)$ is a hitherto unknown function, the metric in EF coordinates can be rewritten as
\begin{eqnarray}
ds^2=\frac{1}{z^2}\left(-f(v,z)g(t,z)^2dt^2+\frac{dz^2}{f(v,z)}+dx_i^2\right) \ .
\end{eqnarray}
The function $g(t,z)$ can now be identified with the redshift factor. In the thin-shell limit we can ignore the compression of the shell as indicated in \cite{yang} and therefore separate the space-time into the following two regimes: exterior and interior. For the exterior geometry, $g(t,z<z_0)=1$ where $z_0$ denotes the position of the shell. This is because the exterior of the shell is described by an AdS-RN geometry. For the interior of the shell, $g(t,z>z_0)=f(v>v_0,z_0)$, where $f(v,z)$ in different dimensions are shown below (\ref{efvaidya4d}) and (\ref{efvaidya5d}). We will refer to this as the quasi-AdS (qAdS) background. For the self-consistency of the paper, we will repeat the computation for the redshift factor in the thin-shell limit shown in 
\cite{yang}. 
From (\ref{dv}), when $v(t,z)=v_c=const$ for $v_c$ representing the center of the shell, the $t$ coordinate will be a function of $z$ even though the function $g(t,z)$ is unknown;
hence the $t$ coordinate can be written as a function of the position of the center of the shell in the $z$ coordinate, denoted by $z_0$. More explicitly, we have
\begin{eqnarray}\label{dz0dt}
\left(\frac{\partial z_0}{\partial t}\right)_{v=v_c}=g(t,z_0)f(v_c,z_0).
\end{eqnarray}
For an arbitrary $z_0$, $v(t(z_0),z=z_0)=v_c$ must be satisfied and fixing $z_0$ is equivalent to fixing $t$. At the thin-shell limit, we may set $v_c=v_0=0$. By revisiting ($\ref{dv}$) at fixed $z_0$ (fixed $t$), we can write down the differential equation encoding the $z$ dependence of $v(t,z)$,
\begin{eqnarray}\label{dvdz}
\left(\frac{\partial v(t(z_0),z)}{\partial z}\right)_{z_0}=-\frac{1}{f(v,z)},
\end{eqnarray}
where we use the center of the shell to represent the position of the shell. Since the mixed derivatives of $v$ with respect to $t$ and $z$ should commute, we have
\begin{eqnarray}\label{dgdz}
\left(\frac{\partial g(t,z)}{\partial z}\right)_t=\left(\frac{\partial f(v,z)}{\partial t}\right)_z\frac{1}{f(v,z)^2}=\left(\frac{\partial f(v,z)}{\partial v}\right)_z\frac{g(t,z)}{f(v,z)^2},
\end{eqnarray}
where we use the chain rule $\left(\frac{\partial f(v,z)}{\partial t}\right)_z=\left(\frac{\partial f(v,z)}{\partial v}\right)_z \left(\frac{\partial v(t,z)}{\partial t}\right)_z=\left(\frac{\partial f(v,z)}{\partial v}\right)_z g(t,z)$ to derive the second equality. We may rewrite (\ref{dgdz}) into integral form by taking the integration from the boundary to the position $z_p$ at fixed $t(z_0)$,
\begin{eqnarray}\label{gz}
g(t(z_0),z_p)=g_0(z_0)\exp\left[\int^{z_p}_{0}\frac{dz}{f(v(t(z_0),z),z)^2}\left(\frac{\partial f(v,z)}{\partial v}\right)_z\right],
\end{eqnarray} 
where $g_0(z_0)=g(t(z_0),0)$ represents the redshift factor on the boundary. This integral can be computed by inserting the solution of $v(t(z_0),z)$ at fixed $t(z_0)$ in (\ref{dvdz}), which can be rewritten as
\begin{eqnarray}\label{gzs}
g(t(z_0),z_p)=g_0(z_0)\exp\left[-\int^{f(v,z_p)}_{f(v,0)}\frac{df(v,z)}{f(v,z)}\right].
\end{eqnarray} 
For the region outside the shell, the spacetime is governed by the AdS-RN metric, which imposes $g_0(z_0)=1$. The redshift factor of the region inside the shell now can be directly obtained by using (\ref{blackeningfun}).
Thus we get:
\begin{eqnarray}\label{metric}
ds^2=\left\{
\begin{array}{ll}
\frac{1}{z^2}\left(-F(z)dt^2+\frac{dz^2}{F(z)}+dx^2\right) & \mbox{if } v>0 \, (z < z_0) \ , \\
\\
\frac{1}{z^2}\left(-F(z_0)^2dt^2+dz^2+dx^2\right) & \mbox{if } v<0 \, (z>z_0) \ ,
\end{array} \right.
\end{eqnarray} 
where 
\begin{eqnarray}\label{Fz}
F(z)=\left\{
\begin{array}{ll}
1-Mz^3+(1/2)M^{4/3}Q^2z^4 \quad {\rm for} \quad  d=3 \ , \\
\\
1-Mz^4+(2/3)M^{3/2}Q^2z^6 \quad {\rm for} \quad d=4 \ .
\end{array} \right.
\end{eqnarray} 

\begin{figure}[h]
\begin{minipage}{8.5cm}
\begin{center}
{\includegraphics[width=8.5cm,height=5cm,clip]{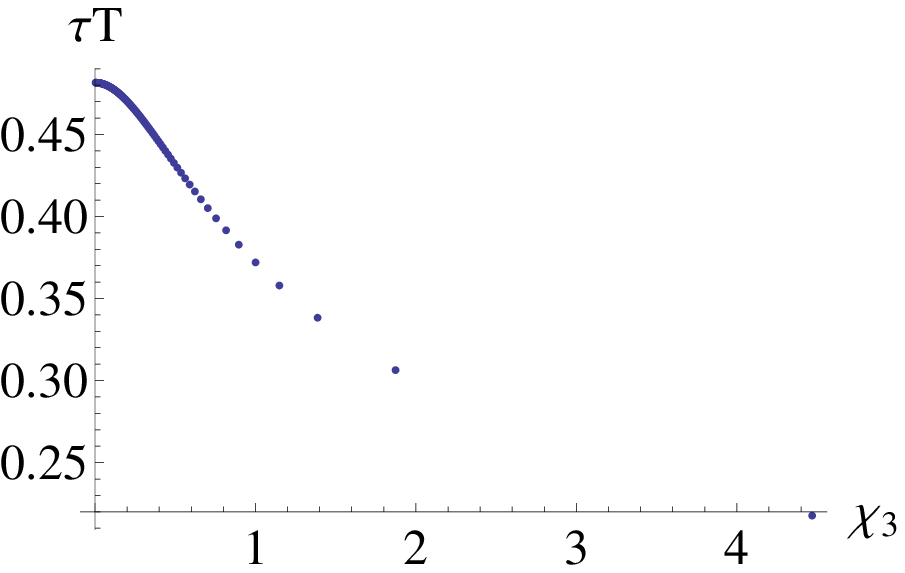}}
\caption{The thermalization time $\tau$ with different values of chemical potential in $d=3$.}\label{thermaltime4d}
\end{center}
\end{minipage}
\begin{minipage}{8.5cm}
\begin{center}
{\includegraphics[width=8.5cm,height=5cm,clip]{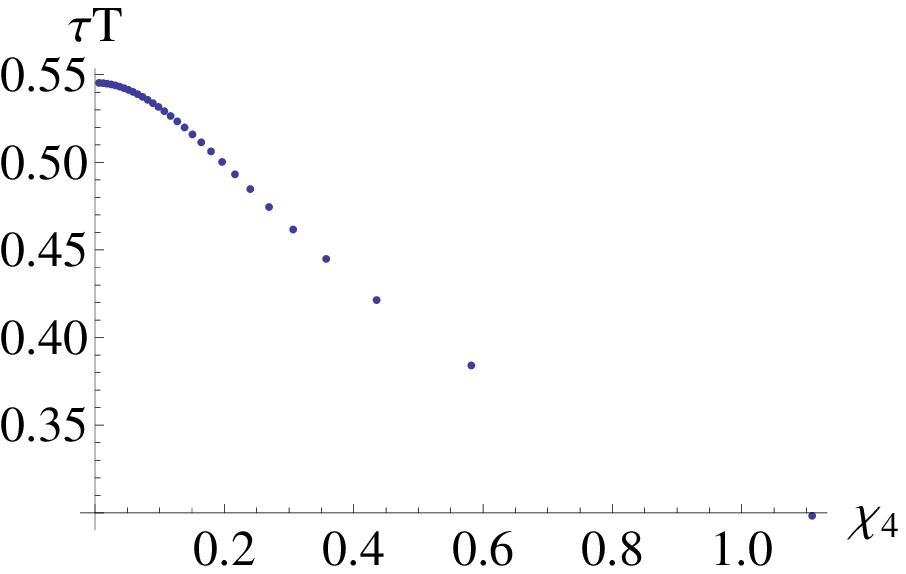}}
\caption{The thermalization time $\tau$ with different values of chemical potential in $d=4$.}\label{thermaltime5d}
\end{center}
\end{minipage}
\end{figure}

Since the falling velocities of the upper and lower surfaces are the same in the thin-shell limit, the position of the falling shell is given by 
\begin{eqnarray}\label{tshell}
t(z_0)=\int^{z_0}_0dz_s\frac{1}{F(z_s)} \ .
\end{eqnarray} 
Given that the shell never coincides with the future horizon exactly in the Poincare coordinates, we approximate the thermalization time as $\tau=t(z_0)|_{z_0=0.99z_h}$, which is shown in Fig.\ref{thermaltime4d} and Fig.\ref{thermaltime5d} with different values of chemical potential measured in the unit of temperature for $d=3$ and $d=4$, respectively. When increasing the chemical potential, the thermalization time decreases. We should comment here that the specific choice of $z_0 = 0.99 z_h$ does not affect the qualitative feature as far as the thermalization time is concerned, any other choice yields a similar behavior. Also, for very large value of $\chi_d$ the thermalization time vanishes asymptotically.

The qualitative behaviors here are distinct from those found by studying non-local operators in the AdS-RN-Vaidya metric\cite{Caceres:2012em}. However, we should notice that the definitions of the thermalization time in these two approaches are different. For example, if we introduce a spacelike geodesic with two ends fixed on the boundary, the thermalization time can be defined as the time when the shell grazes the bottom of the geodesic in the bulk. After that, the whole geodesic will reside in the AdS-RN spacetime, which matches the result derived in the thermalized medium. Therefore, in this situation, the thermalization time of non-local operators are recorded by the position of the falling shell. Nevertheless, when the separation on the boundary of the spacelike geodesic is too large such that the bottom of it penetrates the future horizon, the geodesic can never fully reside in the AdS-RN spacetime even when the shell asymptotically reaches the future horizon. In other words, the maximum thermalization time in Poincare coordinates should be the time when the shell almost reaches the future horizon. For the non-local operators with much larger lengths can never thermalize in Poincare coordinates or the definition of the thermalization time in such a case should be modified. Thus, thermalization times extracted from two approaches can only be compared when the length scale of non-local operators introduced in EF coordinates is about the size of the future horizon. 

\section{\label{sec:level1}The Jet Quenching of Virtual Gluons}

Let us also study thermalization of light probes traversing the non-equilibrium medium. 
In this section, we will follow the approach in \cite{Gubser:2008as} to compute the maximum stopping distance led by the falling massless particle characterizing the tip of a double string with two ends fixed at an infrared scale in the 5-dimensional AdS-RN-Vaidya spacetime. In the gauge theory dual, this double string corresponds to a virtual gluon, in which its energy is encoded by the position of the tip of the string. The computation in the 4-dimensional spacetime can be carried out in the same manner. A similar computation investigating the stopping distance of a massless particle propagating in the bulk as the holographic dual of a R-charged current diffusing on the boundary and the falling string scenario in the AdS-Vaidya spacetime is shown in \cite{yang}.

\subsection{\label{sec:level1}Stopping distance in the AdS-RN spacetime} 

\begin{figure}[h]
\begin{minipage}{8.5cm}
\begin{center}
{\includegraphics[width=8.5cm,height=5cm,clip]{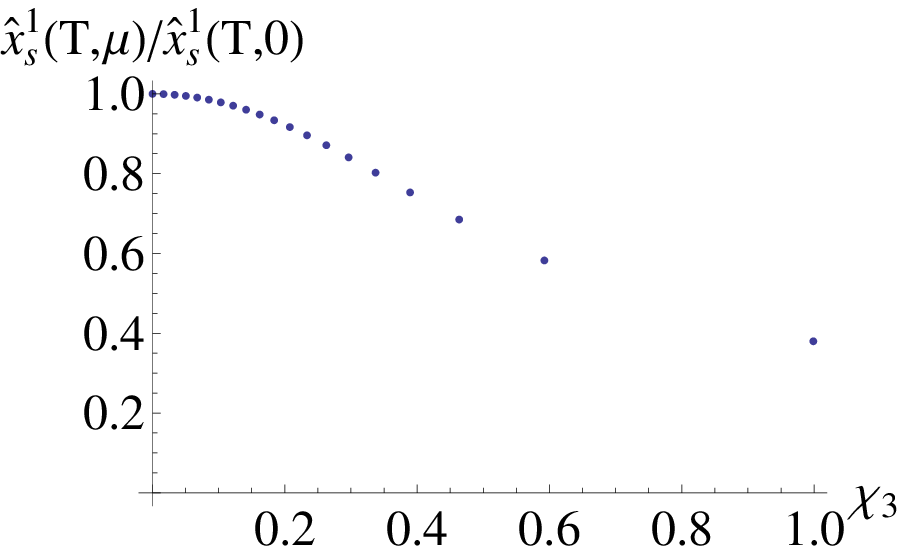}}
\caption{The ratio to the stopping distances with and without chemical potential in the unit of temperature for $d=3$, where $\hat{x}^1_s=x^1_sT$. Here we set $M=1$, $z_I=0$, and $|\vec{q}|=0.99\omega$.}\label{xsratio4d}
\end{center}
\end{minipage}
\begin{minipage}{8.5cm}
\begin{center}
{\includegraphics[width=8.5cm,height=5cm,clip]{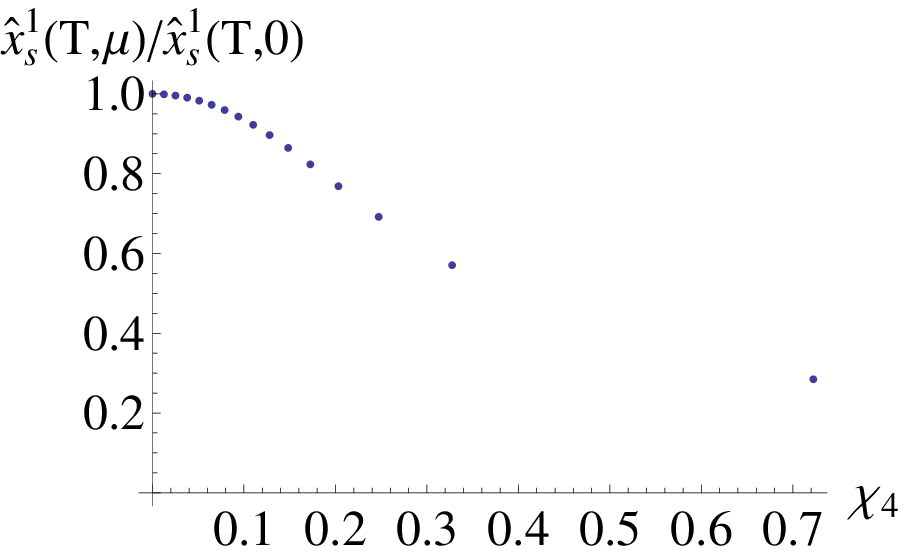}}
\caption{The ratio to the stopping distances with and without chemical potential in the unit of temperature for $d=4$, where $\hat{x}^1_s=x^1_sT$. Here we set $M=1$, $z_I=0$, and $|\vec{q}|=0.99\omega$.}\label{xsratio5d}
\end{center}
\end{minipage}
\end{figure}

Before proceeding with the non-equilibrium medium, we may begin by investigating the thermal medium with a non-zero chemical potential, which is described by the AdS-RN geometry. When the particle moves in the spacetime metric which only depends on the $z$ coordinate, the energy and spatial momentum of the particle should be conserved. The null geodesic can thus be written as \cite{Arnold:2011qi}
\begin{eqnarray}\label{geodesic}
\frac{dx^i}{dz}=\sqrt{g_{zz}}\frac{g^{ij}q_{j}}{(-q_{k}q_{l}g^{kl})^{1/2}} \ ,
\end{eqnarray}
where $g_{ij}$ is the spacetime metric and $q_{i}=(-\omega,|\vec{q}|,0,0)$ for $i=0,1,2,3$ is defined as the conserved 4-momentum of the particle. Here we define the momentum of the particle by using $q_{\mu}=g_{\mu\nu}dx^{\nu}/d\lambda$, $\lambda$ being the affine parameter, in which case we can write down $q_z$ in terms of $q_i$ and $z$. In this note, we will use Roman indices and Greek indices to represent 4-dimensional and 5-dimensional spacetime directions, respectively, unless otherwise mentioned. In the AdS-RN spacetime, (\ref{geodesic}) leads to
\begin{eqnarray}\label{AdSSS}
\nonumber
\frac{dx^1}{dz}&=&\frac{1}{\left(\frac{\omega^2}{|\vec{q}|^2}-F(z)\right)^{1/2}} \ , \\
\frac{dt}{dz}&=&\frac{1}{F(z)\left(1-F(z)\frac{|\vec{q}|^2}{\omega^2}\right)^{1/2}} \ .
\end{eqnarray}
The first equation in (\ref{AdSSS}) allows us to study the stopping distance of the light probe in the thermalized medium with a non-zero chemical potential. The stopping distance is given by
\begin{eqnarray}
x^1_s=\int^{z_h}_{z_I}\frac{dz}{\left(\frac{\omega^2}{|\vec{q}|^2}-F(z)\right)^{1/2}} \ ,
\end{eqnarray}
which only depends on the ratio $|\vec{q}|/\omega$ and the initial position of the tip of the string $z_I$. The ratio $|\vec{q}|/\omega$ roughly represents the ratio to the spatial momentum and energy of the virtual gluon and $z_I$ can be associated with its initial energy. More comprehensive discussions over the physical interpretation of these two quantities can be found in \cite{yang}.   
Even though determining the ratio of $|\vec{q}|$ to $\omega$ entails the initial string profile close to the tip, which could depend on the geometry of the thermalized medium, we may choose the straight string as the simplest setup As shown in \cite{Gubser:2006bz,Herzog:2006gh}, the profile of a moving string obtained from extremizing the Nambu-Goto action in the AdS geometry is a straight string. Although the physical solution in the AdS-Schwarzschild geometry corresponds to a trailing profile, where the profile in the presence of a chemical potential could be more complicated, we will make the same straight-string setup in the thermalized case for comparison. This assures that the gluons traveling in the thermalized medium with different values of chemical potential carry the same initial energy and momentum and as well the same ratio to $|\vec{q}|$ and $\omega$. The ratio of the stopping distance with a non-zero chemical potential to that with zero chemical potential at the same temperature is shown in Fig.\ref{xsratio4d} and Fig.\ref{xsratio5d} for both $d=3$ and $d=4$. It turns out that the stopping distance of the light probe decreases when the chemical potential increases, which has the same qualitative feature compared to the thermalization of the medium.

\begin{figure}[h]
\begin{minipage}{8.5cm}
\begin{center}
{\includegraphics[width=8.5cm,height=5cm,clip]{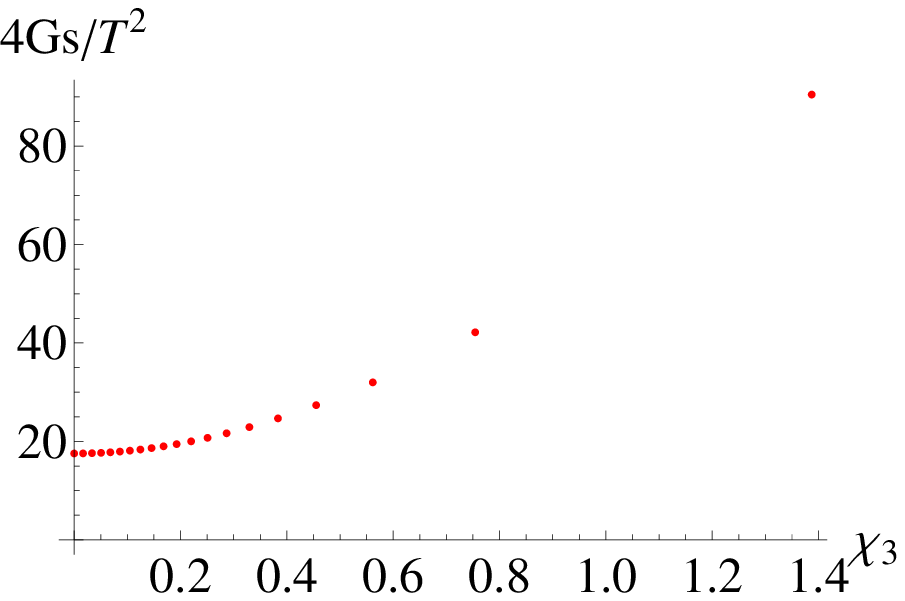}}
\caption{The entropy density with different values of the chemical potential for $d=3$. Here we set $M=1$.}\label{entropy4d}
\end{center}
\end{minipage}
\begin{minipage}{8.5cm}
\begin{center}
{\includegraphics[width=8.5cm,height=5cm,clip]{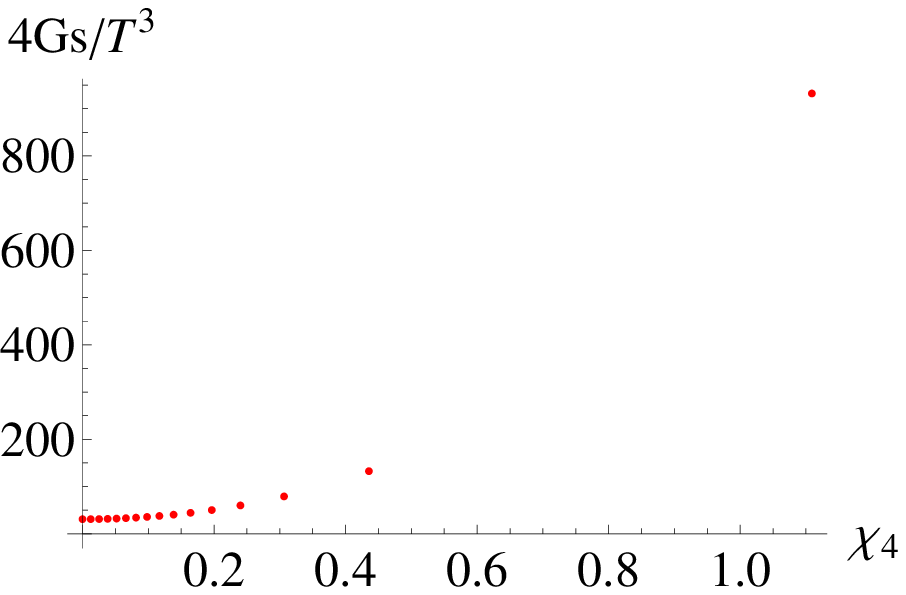}}
\caption{The entropy density with different values of the chemical potential for $d=4$. Here we set $M=1$.}\label{entropy5d}
\end{center}
\end{minipage}
\end{figure}

Physically, by increasing the chemical potential, we actually increase the number of states. From (\ref{entropy}), the entropy density of the plasma in $d+1$ dimension is given by $4Gs=z_h^{-(d-1)}$. As illustrated in Fig.\ref{entropy4d} and Fig.\ref{entropy5d}, the entropy density increases when the chemical potential is increased. 
The medium thus becomes denser, which results in the enhanced scattering for the light probe and a smaller stopping distance. On the other hand, the medium also thermalizes faster because of the same effect. Although the holographic correspondence of the falling shell in the gauge theory side is unknown, the falling shell could be characterized by the collective motion of massless particles with large virtuality in the gravity dual since a null shell is homogeneous along spatial directions and it falls along a null geodesic. Each particle as a component of the shell may as well be influenced by the backreaction to the spacetime metric caused by the falling shell; hence it should qualitatively behave in the same manner as the light probe traversing the medium. Since the component particles of the shell carry large virtuality, the enhanced scattering led by increasing chemical potential would be more pronounced, which reduces the thermalization time of the medium.       
By comparing Fig.\ref{entropy4d} with Fig.\ref{entropy5d}, the entropy density for $d=4$ increases more rapidly than that for $d=3$, which also manifests the steeper drop of the thermalization time of both the probe and the medium for $d=4$ when increasing the chemical potential by comparing Fig.\ref{thermaltime4d} with Fig.\ref{thermaltime5d} and Fig.\ref{xsratio4d} with Fig.\ref{xsratio5d}.

To avoid a naked singularity in the AdS-RN spacetime, there exists a maximum value of $Q$ such that the position of the horizon is given by $F(z) = 0$. By taking $M=1$, the maximum values are $Q_{\rm max}=(27/32)^{1/6}$ and $Q_{\rm max}=3^{-1/4}$ for $d=3$ and $d=4$, respectively. Given that the derivative of $F(z)$ vanishes at the horizon, the temperature of the medium then vanishes. The corresponding zero temperature black hole is known as the extremal black hole. However, this does not correspond to the vacuum since it carries a non-zero entropy given by a non-vanishing area of the extremal black hole horizon. In this situation, we see that the dimensionless parameter $\chi_{4(3)}$ diverges. Also, the stopping distances with a non-zero chemical potential in the unit of temperature, $\hat{x}^1_s(T,\mu)$ shown in Fig.\ref{xsratio4d} and Fig.\ref{xsratio5d}, become zero; we hence lose the ability to make a comparison between the observables in the media with and without a chemical potential in this particular situation.   

\subsection{\label{sec:level1}Initial conditions of the falling string and the matching condition} 

Now we discuss the physics when the system is not in equilibrium. In the qAdS spacetime, energy conservation no longer holds since the redshift factor encoded in the spacetime metric depends on the position of the falling shell. However, by redefining $d\tilde{t}=F(z_0)dt$, the spacetime metric is independent of the new time coordinate $\tilde{t}$. By taking the conserved 4-momentum inside the shell $\tilde{q}_{i}=(-\tilde{\omega},|\vec{q}|,0,0)$, the null geodesic in the quasi-AdS spacetime can be written as
\begin{eqnarray}\label{qAdS}
\nonumber
\frac{dx^1}{dz}&=&\frac{1}{\left(\frac{\tilde{\omega}^2}{|\vec{q}|^2}-1\right)^{1/2}} \ , \\
\frac{d\tilde{t}}{dz}&=&\frac{1}{\left(1-\frac{|\vec{q}|^2}{\tilde{\omega}^2}\right)^{1/2}} \ .
\end{eqnarray}
Now $\omega$ and $\tilde{\omega}$ are two conserved energies defined in the AdS-RN spacetime and the quasi-AdS spacetime, respectively.

When the string fully resides in the quasi-AdS spacetime, the string profile obtained from extremizing the Nambu-Goto action by choosing the redshift time coordinate $\tilde{t}$ and $z$ as the worldsheet coordinates should be a straight string\cite{yang}: $x^1=\tilde{v}\tilde{t}$. The momentum density and momentum of the corresponding string take the form:
\begin{eqnarray}\label{energydensity}\nonumber
\pi^0_{\mu}&=&\frac{\partial\mathcal{L}}{\partial \dot{x}^{\mu}}=(\pi^0_{\tilde{t}},\pi^0_{x^1},\pi^0_{x^2},\pi^0_{x^3},\pi^0_z)=\frac{-1}{2\pi\alpha'z^2\sqrt{1-\tilde{v}^2}}(1,-\tilde{v},0,0,0) \ , \\
p_{\mu}&=&2\int^{\infty}_{z_I}dz\pi^0_{\mu}=\frac{1}{\pi\alpha'z_I\sqrt{1-\tilde{v}^2}}(-1,\tilde{v},0,0,0) \ ,
\end{eqnarray}   
where $\dot{x}^{\mu}=(d x^{\mu}/d \tilde{t})$ and $z_I$ represents the initial position of the tip of the string. It is argued in \cite{yang} that the 4-momentum of the massless particle should be proportional to the 4-momentum of the string since the falling string here behaves approximately as a null string. We thus have the following relation,
\begin{eqnarray}\label{qratio}
\frac{\tilde{q}_i}{\tilde{q}_0}=\frac{p_i}{p_0}=-\tilde{v}.
\end{eqnarray} 
As pointed out in \cite{yang}, the part of a falling string outside the shell could be distorted and bears an unknown profile, while the part inside the shell should remain straight. Nevertheless, the detailed structure of the string profile should not affect the maximum stopping distance. 

\begin{figure}[h]
\begin{minipage}{8.5cm}
\begin{center}
{\includegraphics[width=8.5cm,height=5cm,clip]{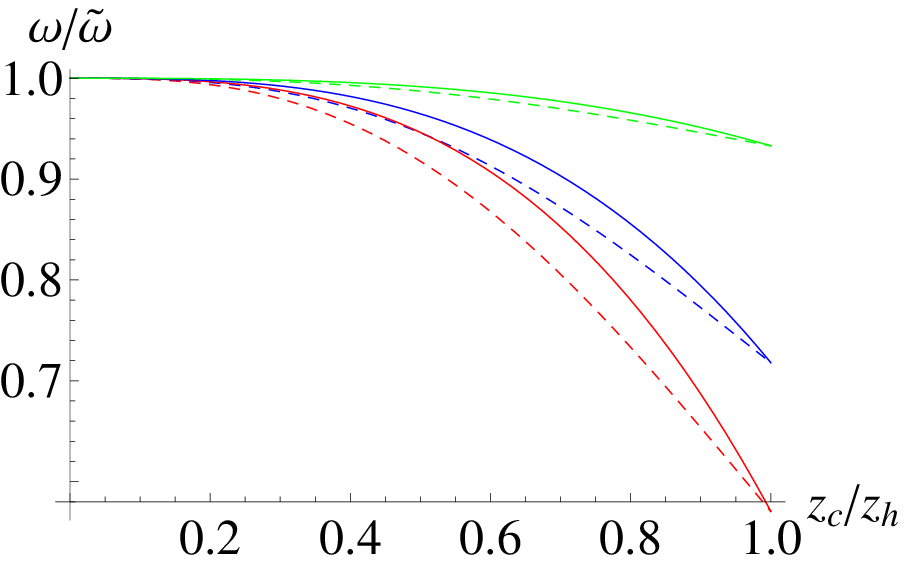}}
\caption{The red, blue, and green solid curves represent the energy ratios for $Q=0$ with $|\vec{q}|=0.99$, $0.9$, and $0.5\tilde{\omega}$, respectively. The case for $Q=0.9$ are showed by the dashed curves. Here we set $M=1$ and $d=3$}\label{omegar4d}
\end{center}
\end{minipage}
\begin{minipage}{8.5cm}
\begin{center}
{\includegraphics[width=8.5cm,height=5cm,clip]{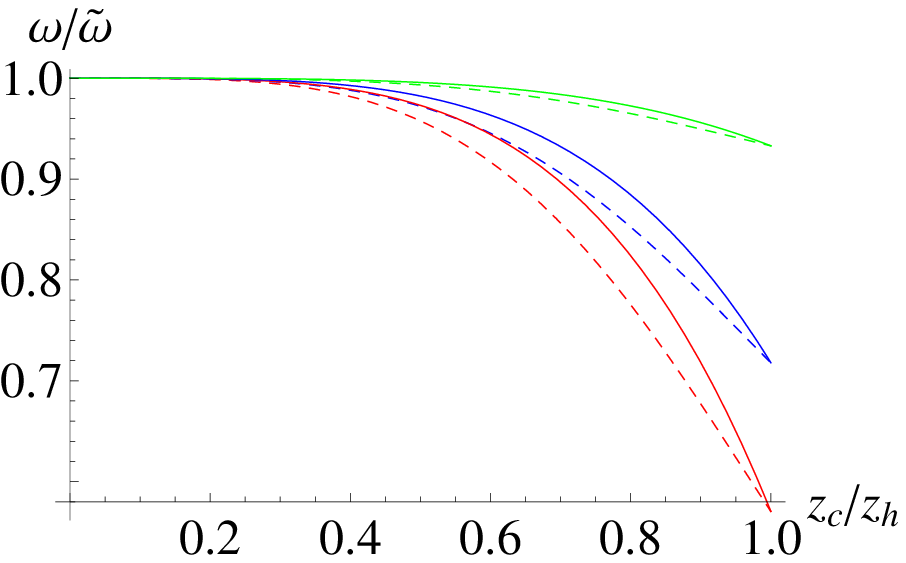}}
\caption{The red, blue, and green solid curves represent the energy ratios for $Q=0$ with $|\vec{q}|=0.99$, $0.9$, and $0.5\tilde{\omega}$, respectively. The case for $Q=0.7$ are showed by the dashed curves. Here we set $M=1$ and $d=4$}\label{omegar5d}
\end{center}
\end{minipage}
\end{figure}

Next, when the tip of the string penetrates the shell and falls in the AdS-RN geometry, we have to find the matching condition to connect the $\tilde{\omega}$ and $\omega$ defined in two spacetimes at the collision point $z_c$. As shown in \cite{yang}, such a condition can be derived by requiring the momentum conservation along the spacetime direction tangent to the shell as an analog of Snell's law. As $p_{\mu}$ and $q_{\mu}$ denote the momenta outside and inside the shell, the matching condition becomes $p_{\mu}V^{\mu}=q_{\mu}V^{\mu}$, where $V^{\mu}=(1,F(z_c),\vec{0})$ represents the tangent vector of the shell. Such a matching condition results in  
\begin{eqnarray}
\frac{\omega}{\tilde{\omega}}=\frac{1}{2(1-\sqrt{1-\delta^2})}[2F(z_c)(1-\sqrt{1-\delta^2})+\delta^2(1-F(z_c))],,
\end{eqnarray}
where $\delta=|\vec{q}|/\tilde{\omega}$.  
The ratios are illustrated in Fig.\ref{omegar4d} and Fig.\ref{omegar5d} for both $d=3$ and $d=4$, respectively. The same matching condition can also be derived in a different setup by using the continuity of the wave function of the massless particle in the WKB approximation \cite{yang}.

\subsection{\label{sec:level1}Stopping distance in the AdS-RN-Vaidya spacetime} 
\begin{figure}[h]
\begin{minipage}{8.5cm}
\begin{center}
{\includegraphics[width=8.5cm,height=5cm,clip]{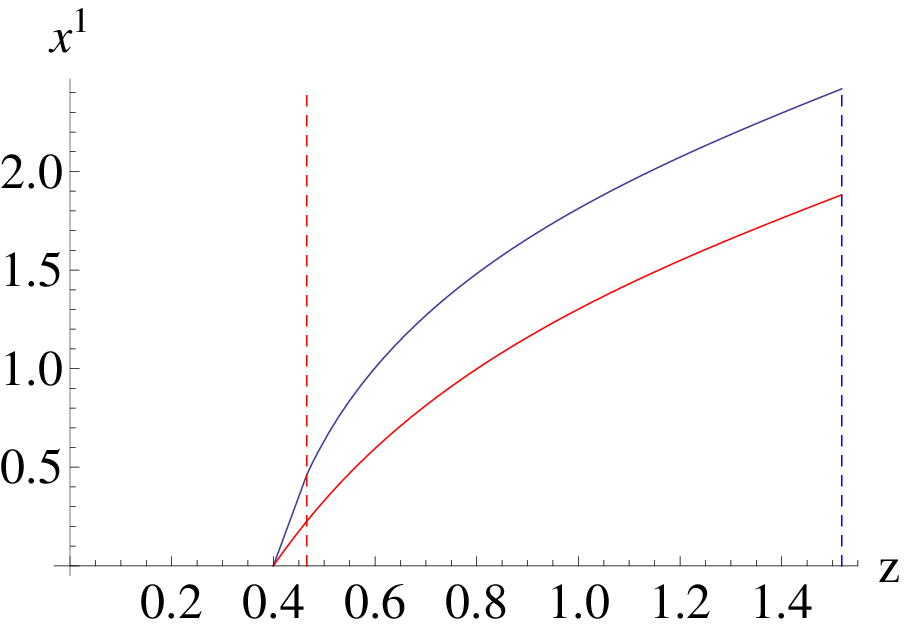}}
\caption{The red and blue curves represent the trajectories of the massless particles moving in AdS-RN and AdS-RN-Vaidya spacetimes for $d=3$ and $\chi_3=4.47$, respectively. The red and blue dashed lines denote the first collision point and the position of the future horizon. Here we take $z_I=0.4$, $M=1$, and $|\vec{q}|/\tilde{\omega}=0.99$ as the initial conditions in both spacetimes.}\label{xz4d}
\end{center}
\end{minipage}
\begin{minipage}{8.5cm}
\begin{center}
{\includegraphics[width=8.5cm,height=5cm,clip]{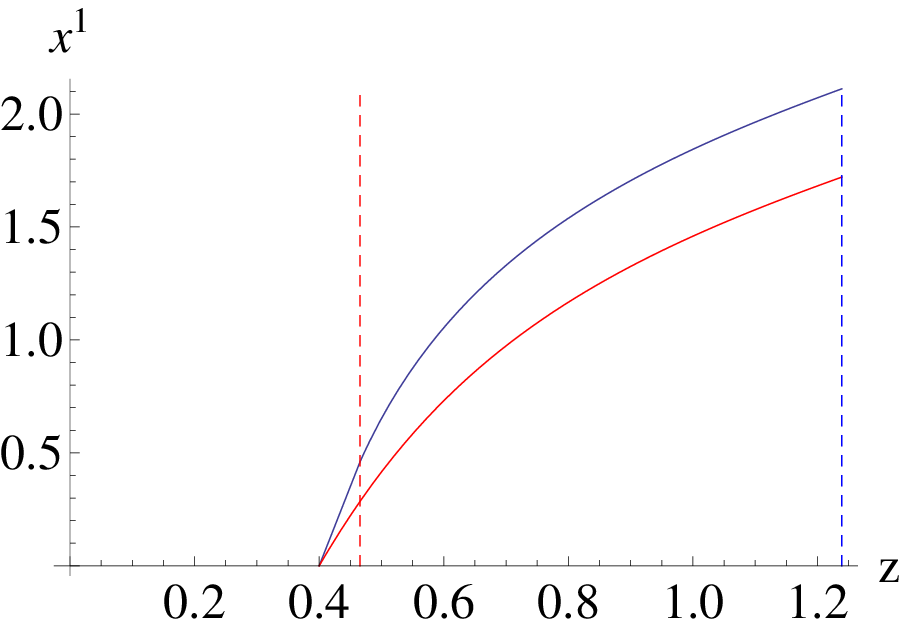}}
\caption{The red and blue curves represent the trajectories of the massless particles moving in AdS-RN and AdS-RN-Vaidya spacetimes for $d=4$ and $\chi_4=1.1$, respectively. The red and blue dashed lines denote the first collision point and the position of the future horizon. Here we take $z_I=0.4$, $M=1$, and $|\vec{q}|/\tilde{\omega}=0.99$ as the initial conditions in both spacetimes.}\label{xz5d}
\end{center}
\end{minipage}
\end{figure}

Now, we may follow the approach in \cite{yang} to compute the stopping distance in the thin-shell limit in Poincare coordinates. We will also eject the shell and massless particle at the same time. The computation involves tracking the shell and the massless particle simultaneously and finding the first collision point $z_c$. For simplicity, here we only show the general expression for the stopping distance, 
\begin{eqnarray}
x^1_s=\int^{z_c}_{z_I}\frac{dz}{\left(\frac{\tilde{\omega}^2}{|\vec{q}|^2}-1\right)^{1/2}}
+\int^{z_b}_{z_c}\frac{dz}{\left(\frac{\omega^2}{|\vec{q}|^2}-F(z)\right)^{1/2}} \ ,
\end{eqnarray}
where the second collision point $z_b$ is approximately equal to $z_h$ since the shell falls with the speed of light.
Similar to the study in the AdS-Vaidya spacetime \cite{yang}, the stopping distance of the massless particle falling from the boundary is not affected by the gravitational collapse in the AdS-RN-Vaidya geometry. In our setup, when we set the initial position of the tip of the string on the boundary, the corresponding virtual gluon on the gauge theory side carries infinite energy as shown in (\ref{energydensity}). As indicated in \cite{yang}, when the hard probe has the energy much larger than any other scale such as the thermalization temperature or the chemical potential of the system, the thermalization of the probe would be insensitive to the thermalization of the medium. 
However, for the soft probe carrying energy comparable to other scales of the system, we may envision an influence of the thermalization process on the jet quenching phenomenon, which is also explored in \cite{yang} for a zero chemical potential. This scenario is shown in Fig.\ref{xz4d} and Fig.\ref{xz5d}, where we eject the massless particle below the boundary, which corresponds to the soft gluon with a finite energy. As a result of thermalization, stopping distance of light probes increases.

\begin{figure}[h]
\begin{minipage}{8.5cm}
\begin{center}
{\includegraphics[width=8.5cm,height=5cm,clip]{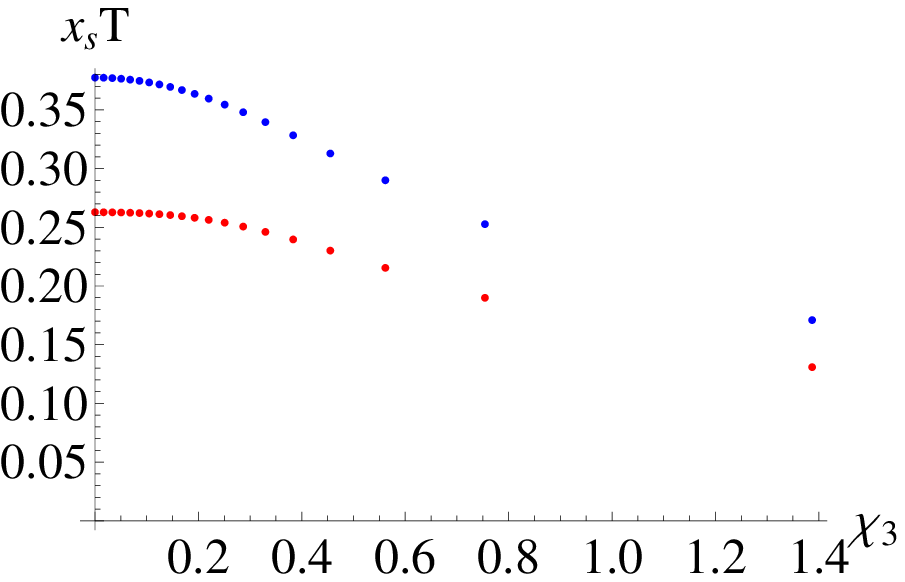}}
\caption{The red and blue points represent the stopping distances with different values of chemical potential in AdS-RN and AdS-RN-Vaidya for $d=3$, respectively. Here we set $M=1$, $z_I=0.4$, and $|\vec{q}|/\tilde{\omega}=0.99$ as the initial conditions in both spacetimes.}\label{xsmu4d}
\end{center}
\end{minipage}
\begin{minipage}{8.5cm}
\begin{center}
{\includegraphics[width=8.5cm,height=5cm,clip]{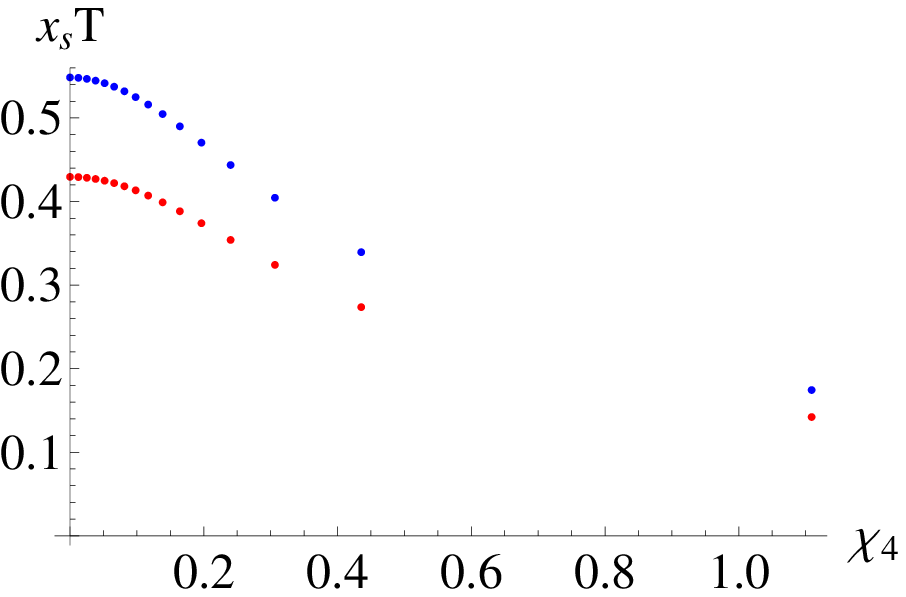}}
\caption{The red and blue points represent the stopping distances with different values of chemical potential in AdS-RN and AdS-RN-Vaidya for $d=4$, respectively. Here we set $M=1$, $z_I=0.4$, and $|\vec{q}|/\tilde{\omega}=0.99$ as the initial conditions in both spacetimes.}\label{xsmu5d}
\end{center}
\end{minipage}
\end{figure}

Finally, we illustrate this behaviour for different values of chemical potential in AdS-RN and AdS-RN-Vaidya spacetimes in Fig.\ref{xsmu4d} and Fig.\ref{xsmu5d}. The results can be reproduced by solving geodesic equations directly in EF coordinates, which are presented in Appendix A; the latter approach can be further applied beyond the thin-shell limit. As shown in Fig.\ref{xsmu4d} and Fig.\ref{xsmu5d}, the stopping distances scaled by the thermalization temperature in both the AdS-RN-Vaidya and AdS-RN geometries decrease when the values of $\chi_d$ are increased and will drop to zero when the values of $\chi_d$ reach infinity.

In general, we find that the probe gluon travels further in the non-equilibrium plasma with a non-zero chemical potential compared to the thermal background. Increasing the magnitude of the chemical potential decreases the stopping distances in both equilibrium and non-equilibrium plasmas.

\section{\label{sec:level1}Summary and Discussions}

In this paper we have investigated thermalization of a non-equilibrium plasma with a non-zero chemical potential by tracking a thin shell falling in the AdS-RN-Vaidya spacetime. We found the thermalization time of the medium decreases when the chemical potential is increased. We have also studied the jet quenching of a virtual gluon traversing such a medium by computing the stopping distance of a falling string, in which the tip of the string falls along a null geodesic. In both the thermalized or thermalizing medium with a non-zero chemical potential, the stopping distance of the probe gluon decreases when the chemical potential is increased. On the other hand, for a soft gluon with finite energy comparable to the thermalization temperature of the medium, its stopping distance in the thermalizing medium is larger than that in the thermalized case.

\begin{figure}[h]
\begin{center}
{\includegraphics[width=8.5cm,height=5cm,clip]{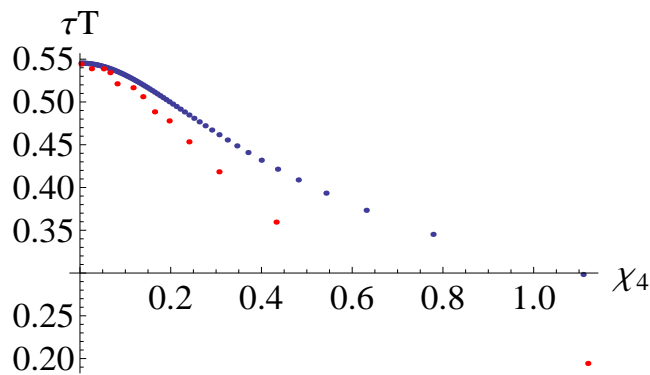}}
\caption{The blue and red points represent the thermalization times scaled by the temperature obtained from our approach and that from analyzing non-local observables, respectively. Here we take $M=1$ in the both cases.}\label{thermaltime}
\end{center}
\end{figure}

In section II, we briefly discussed the difference between the thermalization time obtained from our approach and that derived in \cite{Caceres:2012em}; we will further elaborate on this here. When the medium carries no chemical potential, the position of the horizon is about the inverse of the temperature. As indicated in \cite{yang}, the thermalization times obtained from two approaches in the AdS-Vaidya geometry approximately match when the length scale of the nonlocal operators is about the inverse of the temperature. However, for the medium with a non-zero chemical potential, the temperature does not linearly depend on $z_h^{-1}$. To compare the thermalization times obtained from two approaches, we have to investigate the thermlization time for a nonlocal operator with the length scale about the size of the horizon in the AdS-RN-Vaidya spacetime. As shown in Fig.\ref{thermaltime}, thermalization time obtained by analyzing non-local observables with a length-scale $l_s = 1.71 z_h$\cite{Caceres:2012em} do exhibit similar qualitative feature as the one we have encountered here. Note that the number $l_s/z_h$ bears no possible physical significance other than being an order one number where we have found a visibly pleasant matching. As far as the thermalization time of the medium is concerned, we conclude that our approach by tracking the falling shell close to the horizon seems consistent with that by probing the thermalizing medium with nonlocal operators, when the length scale of the operators approximately equals the size of the horizon. In addition, the thermalization times from both approaches starts to drop more rapidly when $\mu\gtrsim T$.

For the light probe traversing the medium with a non-zero chemical potential,  the decrease of the stopping distance when increasing the chemical potential is expected due to the enhanced scattering with the increasing density of the medium. In \cite{Caceres:2006as,Caceres:2006dj}, it is found that the jet quenching parameter and the drag force of a trailing string in the charged SYM plasma both increase when the chemical potential is increased, which is again consistent with our general observations here.  

\section{\label{sec:level1}Acknowledgement}

The authors thank A. G\"uijosa, B. M\"uller and S. Das for useful discussions. This material is based upon work supported
by DOE grants DE-FG02-05ER41367, de-sc0005396 (D.~L.~Yang), National Science Foundation under Grant Number PHY-0969020 (EC and AK),  CONACyT grant CB-2008-01- 104649 (EC)  and a Simons postdoctoral fellowship awarded by the Simons Foundation (AK).

\section{\label{sec:level1}Appendix}

\subsection{\label{sec:level1}Finding the stopping distance in Eddington-Finkelstein coordinates}

\begin{figure}[h]
\begin{minipage}{8.5cm}
\begin{center}
{\includegraphics[width=8.5cm,height=5cm,clip]{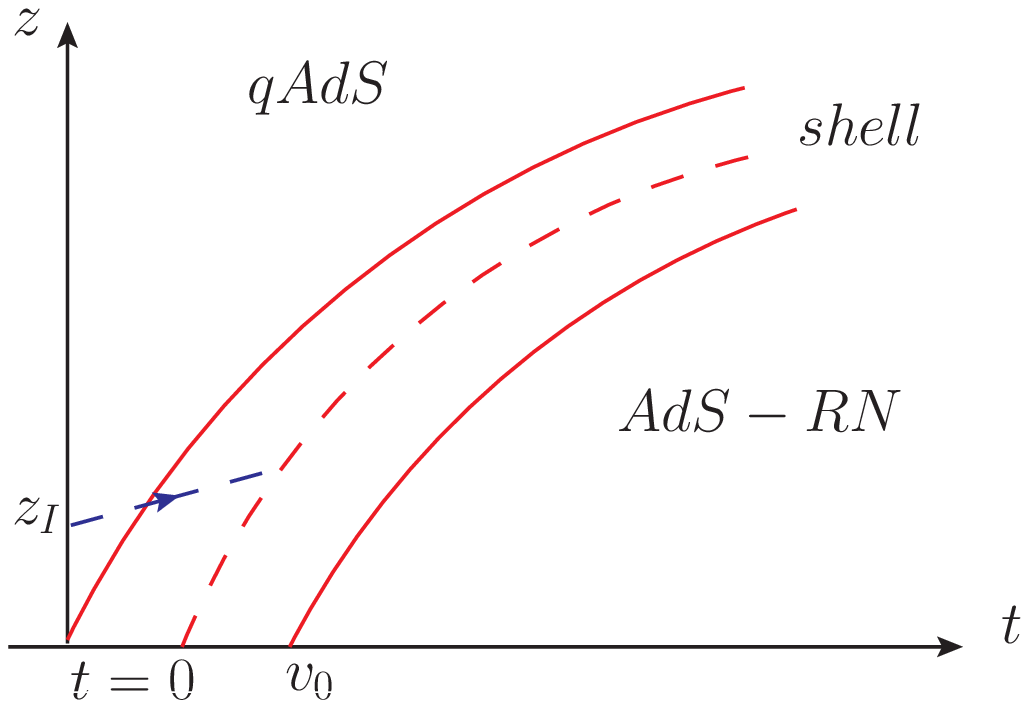}}
\caption{The scenario of a massless particle ejecting from the boundary as the shell starts to fall, where $v_0$ denotes the thickness of the shell and the solid red curves and the dashed red curve represent the surfaces and the center of the shell, respectively. Here the blue dashed line represents the masselss particle ejected from $z_I$.}\label{setup}
\end{center}
\end{minipage}
\begin{minipage}{8.5cm}
\begin{center}
{\includegraphics[width=8.5cm,height=5cm,clip]{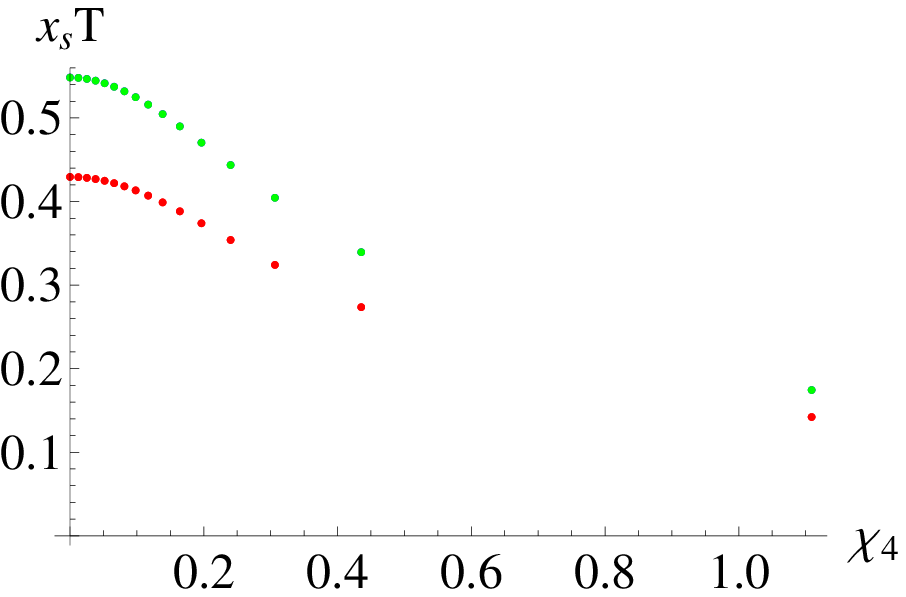}}
\caption{The green points represent the stopping distances in AdS-RN-Vaidya spacetime in EF coordinates, which match those derived in Poincare coordinates as shown by the blue points. Here the initial conditions are the same as those in Fig.\ref{xsmu5d} and we take $v_0=0.0001$.}\label{xsEF}
\end{center}
\end{minipage}
\end{figure}

It is technically difficult to track the null geodesic of the massless particle beyond the thin shell limit in Poincare coordinates, since the trajectory of the particle is continuously deflected by interactions with the shell. However, irrespective of the details of the interaction, it is easier to compute the stopping distance of the probe in EF coordinates. A similar computation in the AdS-Vaidya geometry has been done in \cite{yang}. By using the metric in (\ref{efvaidya4d}) or (\ref{efvaidya5d}) and setting the momentum component of the massless particle non-zero only along $v$, $z$, and $x^1$ directions, we can write down the geodesic equations in terms of the affine parameter $\lambda$,
\begin{eqnarray}\label{geo}\nonumber
\frac{d^2v}{d\lambda^2}&+&\Gamma^v_{vv}\left(\frac{dv}{d\lambda}\right)^2+\Gamma^v_{11}\left(\frac{dx^1}{d\lambda}\right)^2=0\\
\nonumber\frac{d^2z}{d\lambda^2}&+&\Gamma^z_{vv}\left(\frac{dv}{d\lambda}\right)^2
+2\Gamma^z_{vz}\left(\frac{dv}{d\lambda}\right)\left(\frac{dz}{d\lambda}\right)+\Gamma^z_{11}\left(\frac{dx^1}{d\lambda}\right)^2+\Gamma^z_{zz}\left(\frac{dz}{d\lambda}\right)^2=0 \ ,\\
\frac{d^2x^1}{d\lambda^2}&+&2\Gamma^1_{z1}\left(\frac{dz}{d\lambda}\right)\left(\frac{dx^1}{d\lambda}\right)=0 \ .
\end{eqnarray}
For $d=3$, the relevant Chiristoffel symbols are given by
\begin{eqnarray}\nonumber
\Gamma^v_{vv}&=&\frac{1}{z}\left(1+m(v)z^3-\frac{1}{2}Q^2m(v)^{4/3}z^4\right),~\Gamma^v_{11}=-\frac{1}{z} \ , \\
\nonumber\Gamma^z_{vv}&=&\frac{1}{z}\left(-1+\frac{1}{2}m(v)z^3+\frac{1}{2}m(v)^2z^6-\frac{3}{4}Q^2m(v)^{7/3}z^7
+\frac{1}{4}z^8Q^4m(v)^{8/3}\right)\\
&&-\frac{1}{z}\left(\frac{1}{2}z^4-\frac{1}{3}Q^2m(v)^{1/3}z^5\right)\frac{dm(v)}{dv} \ , \\
\nonumber\Gamma^z_{vz}&=&-\frac{1}{z}\left(1+m(v)z^3-\frac{1}{2}Q^2m(v)^{4/3}z^4\right),~  \Gamma^z_{zz}=-\frac{2}{z} \ , \\
\nonumber\Gamma^z_{11}&=&\frac{1}{z}\left(1-m(v)z^3-\frac{1}{2}Q^2m(v)^{4/3}z^4\right),~\Gamma^1_{z1}=-\frac{1}{z} \ .
\end{eqnarray} 
For $d=4$,
\begin{eqnarray}\nonumber
\Gamma^v_{vv}&=&\frac{1}{z}\left(1+m(v)z^4-\frac{4}{3}Q^2m(v)^{3/2}z^6\right),~ \Gamma^v_{11}=-\frac{1}{z} \ , \\
\nonumber\Gamma^z_{vv}&=&\frac{1}{z}\left(-1+m(v)^2z^8+\frac{2}{3}Q^2m(v)^{3/2}z^6
-2Q^2m(v)^{5/2}z^{10}+\frac{8}{9}Q^4m(v)^3z^{12}\right)\\
&&-\frac{1}{2z}\left(z^5-Q^2z^7m(v)^{1/2}\right)\frac{dm(v)}{dv} \ , \\
\nonumber\Gamma^z_{vz}&=&-\frac{1}{z}\left(1+m(v)z^4-\frac{4}{3}Q^2m(v)^{3/2}z^6\right),\mbox{  }\Gamma^z_{zz}=-\frac{2}{z} \ , \\
\nonumber\Gamma^z_{11}&=&\frac{1}{z}\left(1-m(v)z^4+\frac{2}{3}Q^2m(v)^{3/2}z^6\right),\mbox{  }\Gamma^1_{z1}=-\frac{1}{z} \ .
\end{eqnarray} 
In addition to these, the mass function is defined as
\begin{eqnarray}
m(v)=\frac{M}{2}\left(1+\tanh\left(\frac{v}{v_0}-\frac{1}{2}\right)\right) \ ,
\end{eqnarray} 
where the shift in the hyperbolic-tangent function is to fit the setup illustrated in Fig.\ref{setup}. In the thin-shell limit, the mass function defined here will degenerate to the expression in (\ref{mass}). The momentum along the $x^1$ direction $q_1=|\vec{q}|=z^{-2}\frac{dx^1}{d\lambda}$ from the definition $q_{\mu}=g_{\mu\nu}\frac{dx^{\nu}}{d\lambda}$ is conserved in EF coordinates. Thus, we may rewrite the geodesic equations in terms of the derivative with respect to $x^1$,
\begin{eqnarray}\label{derivative}
\nonumber\frac{dx^{\mu}}{d\lambda}&=&\frac{dx^{\mu}}{dx^1}\frac{dx^1}{d\lambda}=z^2|\vec{q}|x'^{\mu} \ , \\
\frac{d^2x^{\mu}}{d\lambda^2}&=&|\vec{q}|^2z^2(x''^{\mu}z^2+2zz'x'^{\mu}) \ ,
\end{eqnarray}
where the prime denotes the derivative with respect to $x^1$. By inserting (\ref{derivative}) into the geodesic equations, we find the last equation in (\ref{geo}) is automatically satisfied due to the conservation of momentum along $x^1$. We then have to introduce proper initial conditions to solve other two geodesic equations in (\ref{geo}). Even though the energy of the particle is not conserved in the AdS-RN-Vaidya spacetime, initially the gravitational effect led by the falling shell is negligible. Therefore, we may write down the initial conditions as a massless particle falling in the pure AdS metric. For a particle ejected from $z_I$, we have
\begin{eqnarray}\nonumber
z|_{x^1=0}&=&z_I \ , ~v|_{x^1=0}=t|_{x^1=0}-z|_{x^1=0}=-z_I \ , \\
\nonumber v'|_{x^1=0}&=&(t'-z')|_{x^1=0}=\frac{\omega_0}{|\vec{q}|}-\frac{\sqrt{\omega_0^2-|\vec{q}|^2}}{|\vec{q}|}
 \ ,\\
z'|_{x^1=0}&=&\frac{\sqrt{\omega_0^2-|\vec{q}|^2}}{|\vec{q}|} \ ,
\end{eqnarray}
where $\omega_0$ and $\vec{q}$ represent the initial energy and momentum of the massless particle in Poincare coordinates. The stopping distance obtained in EF coordinates is shown in Fig.\ref{xsEF} for $d=4$ in the thin-shell limit, where we also make a comparison with the result illustrated in Fig.\ref{xsmu5d} with the same initial conditions. The results match and cannot be distinguished from the plot. This holds for $d=3$ as well. We can also evaluate the stopping distance with the thick shell in EF coordinates, it turns out that the deviation from the thin shell is negligible.  

\subsection{\label{sec:level1}The dyonic black hole}    

Here we will summarize some properties of the dyonic black hole in $d=3$. The action we consider is the following
\begin{eqnarray}
S_0 = \frac{1}{8\pi G_N} \left(\frac{1}{2} \int d^4x \sqrt{-g} ( R - 2 \Lambda ) - \frac{1}{4} \int d^4 x F^2 \right) \ ,
\end{eqnarray}
which results in the following eom
\begin{eqnarray}
&& R_{\mu\nu} - \frac{1}{2} \left( R - 2 \Lambda \right) g_{\mu\nu} = g^{\alpha\rho} F_{\rho\mu} F_{\alpha\nu} - \frac{1}{4} g_{\mu\nu} F^2 \ , \\
&& \partial_\rho \left[ \sqrt{-g} F^{\rho\sigma} \right] = 0 \ , \quad \partial_\rho \left[ \sqrt{-g} (\star F)^{\rho\sigma} \right] = 0 \ ,
\end{eqnarray}
where $(\star F)$ is the Hodge dual of $F$. In $(3+1)$-bulk dimensions, the Hodge dual of the bulk electro-magnetic field is another two form. Thus it is possible for a black hole to possess both electric and magnetic charges. Such a dyonic black hole is given by
\begin{eqnarray}
&& ds^2 = \frac{L^2}{z^2} \left( - f dt^2 + \frac{dz^2}{f} + d\vec{x}^2 \right) \ , \quad \Lambda = - \frac{3}{L^2} \ , \\
&& f(z) = 1 - M z^3 + \frac{1}{2L^2} \left(Q_e^2 + Q_m^2 \right) z^4 \ , \\
&& F = Q_e dz \wedge dt + Q_m dx \wedge dy \ ,
\end{eqnarray}
where $Q_e$ and $Q_m$ denote the electric and the magnetic charges respectively.

In the extremal case, the function $f$ can be written as
\begin{eqnarray}
f(z) = 1 - 4 \left(\frac{z}{z_H}\right)^3 + 3 \left(\frac{z}{z_H}\right)^4 \ ,
\end{eqnarray}
where $z_H$ is the location of the event-horizon. It is easy to check that the temperature is given by
\begin{eqnarray}
4 \pi T = - \left. \frac{df}{dz} \right| _{z_H} = 0 \ .
\end{eqnarray}
Therefore for the extremal case, we can write
\begin{eqnarray}
z_H^3 = \frac{4}{M} \ , \quad Q_e^2 + Q_m^2 = \frac{6}{L^2} \frac{1}{z_H^4} \ .
\end{eqnarray}

We can now find the corresponding Vaidya-type background sourced by appropriate matter field
\begin{eqnarray}
S = S_0 + \kappa S_{\rm ext} \ .
\end{eqnarray}
This will give the following equations of motion
\begin{eqnarray}
&& R_{\mu\nu} - \frac{1}{2} \left( R - 2 \Lambda \right) g_{\mu\nu} - g^{\alpha\rho} F_{\rho\mu} F_{\alpha\nu} + \frac{1}{4} g_{\mu\nu} F^2 = \left(16 \pi G_N \kappa\right) T_{\mu\nu}^{\rm ext}\ , \\
&& \partial_\rho \left[ \sqrt{-g} F^{\rho\sigma} \right] = \left( 8\pi G_N \kappa\right) J_e^{\sigma} \ , \\
&& \partial_\rho \left[ \sqrt{-g} (\star F)^{\rho\sigma} \right] =  \left( 8\pi G_N \kappa\right) J_m^{\sigma} \ .
\end{eqnarray}
Here $J_e$ denotes the electric current and $J_m$ denotes the magnetic current. The dyonic-Vaidya background takes the following form
\begin{eqnarray}
ds^2 = \frac{L^2}{z^2} \left( - f dv^2 - 2 dv dz + d\vec{x}^2 \right) \ , \quad f(z,v) = 1 - m(v) z^3 + \frac{1}{2 L^2} \left(q_e(v)^2 + q_m(v)^2 \right) z^4 \ ,
\end{eqnarray}
with the following vector fields
\begin{eqnarray}
F_{zv} = q_e(v) \ , \quad F_{xy} = q_m(v) \ .
\end{eqnarray}
The above background is sourced by the following stress-energy tensor
\begin{eqnarray}
2\kappa T_{\mu\nu}^{\rm ext} = \frac{z^2}{L^2} \left(L^2 \frac{dm}{dv} - z \left( q_e \frac{dq_e}{dv} + q_m \frac{dq_m}{dv} \right) \right) \delta_{\mu v} \delta_{\nu v} \ .
\end{eqnarray}
The electric and magnetic sources are given by
\begin{eqnarray}
\kappa J_e^\mu = \frac{d q_e}{dv}\delta^{\mu v} \ , \quad \kappa J_m^\mu = \frac{d q_m}{dv}\delta^{\mu v} \ .
\end{eqnarray}
The extremal limit is now obtained by considering
\begin{eqnarray}
z_H(v)^3 = \frac{4}{m(v)} \ , \quad q_e(v)^2 + q_m(v)^2 = \frac{6}{L^2} \frac{1}{z_H(v)^4} \ .
\end{eqnarray}
Once $m(v)$ is chosen, we can pick the electric and magnetic charge functions according to the above formula. An obvious such choice is given by
\begin{eqnarray}
q_e(v)^2 = \frac{3}{4L^2} \left(\frac{m(v)}{\sqrt{2}}\right)^{4/3} = q_m(v)^2 \ .
\end{eqnarray}
%


\end{document}